\documentclass[seceq,supplement]{ptptex}

\usepackage{graphicx}

\title{Democratic Superstring Field Theory and its Gauge Fixing}

\author{Michael \textsc{Kroyter}}

\inst{Tel-Aviv University\\ The Raymond and Beverly Sackler Faculty of Exact Sciences\\
Tel Aviv University, Ramat Aviv, 69978, Israel}

\abst{
This work is my contribution to the proceedings of the conference
``SFT2010 -- the third international conference on string field theory and related topics''
and it reflects my talk there, which described the democratic string field theory and
its gauge fixing.

The democratic string field theory is the only fully RNS string field theory to date.
It lives in the large Hilbert space and includes all picture numbers. Picture changing
amounts in this formalism to a gauge transformation.
We describe the theory and its properties and show that when partially gauge fixed it can be
reduced to the modified theory and to the non-polynomial theory. In the latter case we can even
include the Ramond sector in the picture-fixed action.
We also show that another partial gauge-fixing leads to a new
consistent string field theory at picture number $-1$.}

\newcommand{\La} {\Lambda}
\newcommand{\al} {\alpha}
\newcommand{\cO} {{\cal O}}
\newcommand{\cz} {{\mathbb Z}}
\newcommand{\Ztwo}{{\mathbb{Z}_2}}
\newcommand{\dint} {{\mbox{\reflectbox{${\displaystyle \int}$}}}}

\begin{document}

\maketitle

\section{Introduction}

Could string field theory serve as a non-perturbative definition
of string theory?
An obvious pre-requirement for that to be the case is that the string field theory
should exist\footnote{See~\cite{Fuchs:2008cc} for a review of recent results in string
field theory. In particular, sections 8.1 and 8.2 introduce superstring field theories.}.
Moreover, a string field theory that has the chance of really ``capturing the essence''
of string theory should probably be universal, i.e., it should not depend on any
properties that are peculiar to a given background, but at most on entities that exist
in all backgrounds, e.g., the BRST charge $Q$. A universal string field theory
would in particular be Poincar\'e-covariant in the case of a flat background.

Universal string field theories exist for the bosonic cases:
Witten's open string field theory~\cite{Witten:1986cc}
and Zwiebach's closed string field theory~\cite{Zwiebach:1992ie}.
However, according to the current paradigm, the bosonic theories are
``toy-models'' and the ``physical theories'' are the supersymmetric ones.
Hence, what we need in order to even pose the question above is a
superstring field theory. 

Superstring theories can be described using several formulations.
Most familiar are the Ramond-Neveu-Schwarz (RNS), Green-Schwarz (GS) and
pure-spinor formulations. It is not quite clear how should the GS formalism be
treated covariantly. As for the pure-spinor approach, it is not clear how to describe
the GSO($-$) sectors using it (although some results on this matter
exist~\cite{Berkovits:2007wz}). Hence, it cannot be considered fully universal.
Furthermore, while a pure-spinor string field theory
exists~\cite{Berkovits:2005bt}, there is evidence that the pure-spinor formulation
should be altered in order to obtain a pure-spinor string field theory that is
consistent at the quantum level~\cite{Aisaka:2008vw,Bedoya:2009np}.

In light of all that it seems that the RNS formalism is the most promising one
for the construction of universal superstring field theories. Indeed, that
was the formalism used by Witten, immediately after his construction of the
bosonic theory, in his attempt to construct a superstring field
theory~\cite{Witten:1986qs}. Unfortunately, it was soon realised that
Witten's theory suffers from singularities~\cite{Wendt:1987zh}.

The source of the singularities can be traced to the existence, in the
RNS formalism, of a new quantum number, namely the picture number~\cite{Friedan:1985ge}.
Each physical vertex operator is infinitely degenerate. NS vertex operators
carry integer picture numbers and Ramond vertex operators carry half-integer
picture numbers. In order to get a non-trivial CFT expectation value
the picture numbers of all insertions should sum up to $-2$ if one works
in the small Hilbert space, as in the common description of RNS 
vertex operators and to $-1$, in the case of the large Hilbert space.
When one evaluates expectation values of vertex operators the final
result is independent on the distribution of picture numbers among the
vertex operators, as long as they sum up to the desired result.
Hence, all picture numbers are equivalent in this case and one can even
change the picture of vertex operators using the picture changing operators
to an arbitrary (integer for NS or half-integer for R) value.
On the other hand, for off-shell string fields there is no notion of
picture changing and the string fields at different picture numbers are
generically inequivalent. In his construction, Witten removed the degeneracy
related to picture number by choosing specific values for the NS and R
string fields. Then, he had to introduced explicit insertions of picture
changing operators at the string mid-point, in the action and in the gauge
transformation, in order to obtain a non-zero result and in order to
remain at the desired picture numbers, respectively. The iteration of
(linearised) gauge transformations with this explicit mid-point insertion
is what led to the singularities, due to the invariance of the mid-point
under the star product.

There were several attempts towards an improved RNS open string field theory.
However, until recently, all of them shared the idea of resolving the degeneracy
related to picture number by choosing a
unique picture number for any given string field. This idea is unnatural
from a string field theoretical perspective, since, as we stated,
string fields are generically inequivalent at different picture numbers.
One such construction goes under the name of the modified
theory~\cite{Arefeva:1989cp,Preitschopf:1989fc}. While this theory led
to some impressive results at the NS sector~\cite{Erler:2007xt}, it was
proven that its Ramond sector suffers from the same problems as Witten's
construction~\cite{Kroyter:2009zi}.
Another theory is the non-polynomial NS string field theory of
Berkovits~\cite{Berkovits:1995ab,Berkovits:2001im}, which is usually considered
to be the most reliable NS string field theory we have. This theory avoids explicit
insertions of picture changing operators by being defined at the large
Hilbert space while introducing a new gauge symmetry.
Unfortunately, a universal incorporation of the Ramond sector into this theory can only be
achieved by imposing constraints that cannot be
derived from an action~\cite{Michishita:2004by}.

This situation changed with the introduction of the ``democratic string field
theory''\cite{Kroyter:2009rn}. This theory treats the picture-redundancy as a gauge symmetry,
following the ideas of\cite{Berkovits:2001us}. Thus, the theory is defined using
string fields whose picture is arbitrary, i.e., over the whole ``Hilbert space''.
Moreover, the NS and R string fields are naturally unified within this theory.
Furthermore, partial gauge fixings of the NS sector of this theory lead to
the modified as well as to the non-polynomial theories~\cite{Kroyter:2010rk}.
We introduce the main ingredients of the democratic theory in~\ref{sec:democratic}
and follow to describe the partial gauge fixings in~\ref{sec:gauge}.
Conclusions and outlook are offered in~\ref{sec:conc}.

\section{The democratic theory}
\label{sec:democratic}

String fields are off-shell generalizations of vertex operators. There are several possible
representations of RNS vertex operators and different such representations
would lead to different string field theory formulations. Hence, the first step
towards the construction of an RNS string field theory is to decide upon a representation.

The different representations of RNS vertex operators originate from
the fact that while the natural space in which the RNS string is defined is the large Hilbert space,
the space in which the cohomology problem is most simply formulated is the small
Hilbert space. These spaces are related by
\begin{equation}
H_S=\eta_0 H_L\,,\qquad H_L=H_S\oplus \xi_0 H_S\,,
\end{equation}
where $\eta_0$ and $\xi_0$ are the zero mode of $\eta$ and $\xi$, which together with $\phi$ are the conformal fields
obtained in the process of ``bosonizing'' the $\beta\gamma$ ghost system~\cite{Friedan:1985ge}.

The possible representations of vertex operators, as formulated in the large Hilbert space are
($\Psi$ is the vertex operator, $\delta \Psi$ is the ``gauge transformation'' defining the equivalence within the cohomology
classes and the picture number is integer for NS and half-integer for R vertex operators),
\begin{equation}
\nonumber
\begin{array}{lll}
1.\ \eta_0\Psi = Q\Psi = 0\,,& \delta \Psi=Q \Lambda \qquad \eta_0\Lambda=0 & pic(\Psi)=pic(\Lambda)=p_0\,.\\
2.\ \eta_0 Q \Psi = 0\,, &\delta \Psi=Q\Lambda_Q+\eta_0 \Lambda_\eta\qquad & pic(\Psi)=pic(\Lambda_Q)=pic(\Lambda_\eta)-1=p_0\,.\\
3.\ (Q-\eta_0)\Psi=0\,, &\delta \Psi=(Q-\eta_0)\Lambda & -\infty\leq p_0<pic(\Psi)<p_1\leq \infty\,.
\end{array}
\end{equation}
The inequalities above should be interpreted as meaning that $\Psi$ has components at the specific range of picture numbers.
We see that the first two representations differ from the third one by the fact that they have a fixed picture number.
While in the first two cases the cohomologies that one gets at different picture numbers are equivalent,
taking several picture numbers together leads to an appropriate doubling of the cohomology. In contrast,
the third representation gives the same cohomology (that also matches of course those of the first two cases),
regardless of the range of the pictures. In this case, one can always ``push the cohomology'' to any particular
picture. In the case of an infinite range of picture numbers one has to impose the requirement that the
coefficients obtained by the pushing are absolutely convergent. Working with an infinite domain of
pictures has the added advantage that no restrictions should be put on the gauge transformations,
i.e., when the domain is finite a gauge transformation might bring the vertex operator outside of its
allowed picture domain and the gauge string fields $\Lambda$ should be restricted in order to prevent
that.

The modified theory is a natural string field theoretical generalisation of the first representation
(with $p_0=0$). A generalisation of the second representation (again at $p_0=0$) leads to the
non-polynomial theory. The third representation can also be generalised to a string field theory,
but only in the case of a totally unrestricted range of picture numbers ($p_1=-p_0=\infty$).
It is not quite clear how to generalise the criterion of absolute convergence to the off-shell case.
This is just one more manifestation of the usual problem with defining a space of string
fields\footnote{A collection of some ideas regarding this space can be found in section 2 of~\cite{Kroyter:2009zj}.
Some new, yet unpublished, ideas were presented by Leonardo Rastelli in this conference and in others.}.
Since the theory uses all possible picture numbers it is referred to as the ``democratic'' theory.

The democratic theory is defined by the action (we keep the star product implicit and use the following integration
symbol in order to represent large Hilbert space expectation value)
\begin{equation}
\label{action}
S=-\oint \cO\Big(\frac{1}{2}\Psi(Q-\eta_0)\Psi+\frac{1}{3}\Psi^3\Big),
\end{equation}
where $\Psi$ is the string field that carries, like the vertex operators, ghost number one.
The operator $\cO$ that appears in the action is a mid-point insertion, which is a regularized sum of $\xi$ times the sum
of all multi-picture-changing-operators $X_p$,
\begin{equation}
\label{OandXp}
\cO\approx \xi \sum_{p\in \cz} X_p \qquad \Longleftrightarrow \qquad Q\cO=\eta_0 \cO=\sum_{p\in \cz} X_p\,.
\end{equation}
It was proven in~\cite{Kroyter:2009rn} that these operators
exist for all integer $p$ (in particular $X_0=1$, $X_1=X=Q\xi$ and $X_{-1}=Y=c\partial \xi e^{-2\phi}$),
that they obey the relation,
\begin{equation}
X_{p_1}(z)X_{p_2}(w)\sim X_{p_1+p_2}(w)+Q\eta_0 (\cdots)\,,
\end{equation}
that they are all zero-weight conformal primaries and that the $\cO$ defined
using them is also a zero-weight conformal primary. While these requirements do not fix $\cO$ completely, it was shown,
along the lines of a similar proof for the modified theory~\cite{Kroyter:2009bg}, that theories with
different legitimate choices of $\cO$ are (at least) classically equivalent.

The democratic theory has the following (linearised) gauge invariance
\begin{equation}
\label{gauge}
\delta \Psi=(Q-\eta_0)\Lambda+[\Psi,\Lambda]\,.
\end{equation}
This gauge symmetry is a non-linear generalisation of the one we had in the cohomology problem.
Thus, it includes the usual gauge symmetry together with other gauge transformations that generalise
the notion of picture changing to off-shell string fields.
The fact that the gauge transformation does not include explicit picture changing operators allows us
to restrict the string fields to configurations that do not include mid-point insertions. Hence, it
is impossible to induce the singularities that afflict Witten's theory and the modified theory.

The string field $\Psi$ can include both NS and R sectors and when decomposed as $\Psi=A+\alpha$ with $A$ in the NS
sector and $\alpha$ in the R sector the correct equations for both sectors are obtained.
One can further introduce internal and external Chan-Paton factors and define the theory over an arbitrary
brane-system. Furthermore, the gauge symmetry of the theory, which is infinitely reducible and closes only
on-shell, can be simply treated using
the BV formalism. All that is needed is to relax the constraint of $\Psi$ carrying ghost number one.
The resulting action still looks like~(\ref{action}) and it obeys the classical master equation.

A simple partial gauge fixing of the NS sector of the democratic theory can be obtained by restricting the
string field to live in the small Hilbert space and to carry picture number zero. This partial gauge fixing
results in nothing but the modified theory and the residual gauge symmetry is the usual gauge symmetry of this
theory. This implies that solutions of the modified theory, such as~\cite{Erler:2007xt}
are also solutions of the democratic theory. All the above gives much credibility to the democratic theory.

\section{Gauge fixings}
\label{sec:gauge}

Despite the immediate successes of the democratic theory there are many issues that should be clarified.
A central such issue is that of gauge fixing. One might conclude that it is sensible to gauge fix the
theory to any given picture number, which at picture number $-1$ would presumably lead to Witten's
theory and at other picture numbers to even more erroneous results. Another important question is
the relation between this formalism and the other well-established (NS) formalism, namely the non-polynomial
one\footnote{We already mentioned that the democratic theory can be partially gauge fixed to the modified theory
and it was claimed in~\cite{Fuchs:2008zx} that this theory is equivalent to the
non-polynomial theory. However, this equivalence is defined only for solutions. Moreover, it explicitly
uses the mid-point and as such is susceptible to mid-point singularities.}.
Here, we would like to claim that a generic fixing of the picture number is not allowed and that the
gauge fixing to picture number $-1$ leads not to Witten's theory but to a new consistent (NS) theory.
We further show that another gauge fixing can be used in order to obtain the non-polynomial theory
from the democratic theory. We focus on the NS sector from now on and only comment on the Ramond sector,
for simplicity.

It is clear that at the linearised, vertex operator, level one can fix the picture number to an arbitrary
value, while further restricting the string field to lie in the small Hilbert space.
Indeed, the residual gauge symmetry is then generated by a gauge string field $\Lambda$ with
a well defined picture number. All that is changed upon the inclusion of the non-linear term in~(\ref{gauge}).
Now, the last term in the r.h.s of~(\ref{gauge}) generates string fields with a different picture number,
unless $\La$ carries picture number zero, in which case the resulting string field also carries zero picture
number. The restriction of $\La$ to zero picture number becomes even clearer when one considers the
finite form of the gauge transformation,
\begin{equation}
\Psi\rightarrow e^{-\Lambda}(Q-\eta_0+\Psi)e^\Lambda\,.	
\end{equation}
A generic $\Lambda$ with a well defined picture number would induce infinitely many picture numbers unless
its picture number equals zero. We see that in order to remain with a sensible residual gauge symmetry
the picture number of $\La$ must equal zero. In this case we can either further restrict
$\Psi$ to have $pic(\Psi)=0$ in the small Hilbert space,
\begin{equation}
\label{etaPsi0}
\eta_0 \Psi=0\,,
\end{equation}
or to carry picture number
$-1$ and live in the dual Hilbert space~\cite{Berkovits:1994vy},
\begin{equation}
\label{QPsi0}
Q \Psi=0\,.
\end{equation}

The fact that not all possible picture numbers can be used for gauge fixing originates from the non-linearities
of the gauge transformation. We have ``three'' types of redundancies that we have to fix:
\begin{itemize}
\item The ``usual'' gauge freedom.
\item The two-fold redundancy related to the use of the large Hilbert space.
\item The infinite redundancy of picture numbers.
\end{itemize}
While at the linearised level and when restricting to vertex operators these are three independent
redundancies, the introduction of non-linearities entangles them. Gauge fixing to a given picture
number is, thus, quite unnatural from the perspective of the democratic theory and as we claimed
might only be possible at pictures $0$ and $-1$. One can consider other partial gauge fixings,
e.g., for any $\La$, whose picture number is non-zero, the set $e^{-\La}(Q-\eta_0+\Psi)e^\La$,
with a zero-picture-number $\Psi$ living in the small Hilbert space, can be considered as a partially
gauge fixed space of string fields.

\subsection{Gauge fixing at $pic(\Psi)=0$}

Assume now that we enforce the gauge choice $pic(\Psi)=0$. We should still take care of the
redundancy related to the use of the large Hilbert space. At the level of vertex operators
we have the two conditions~(\ref{etaPsi0}) and~(\ref{QPsi0}) and we can choose either one
of those as an additional gauge condition, in order to obtain the standard form of the
cohomology problem. When the non-linearities are considered we have to check whether
any of these conditions in consistent with the desired form of the residual gauge symmetry.
One can see that the first case~(\ref{etaPsi0}) is indeed consistent. Substitution into
the action leads, as already mentioned, to the modified theory.

The second case~(\ref{QPsi0}) is by itself inconsistent, but has a natural non-linear extension.
The fact that the cohomology of $Q$ in the large Hilbert space is trivial implies
that~(\ref{QPsi0}) is equivalent to
\begin{equation}
\exists \Phi:\quad \Psi=Q \Phi\,.
\end{equation}
We generalise this restriction to,
\begin{equation}
\label{PhiForPsi}
\exists \Phi:\quad \Psi=e^{-\Phi}Q e^\Phi\,.
\end{equation}
Note, that this is {\it not} a restriction to pure-gauge-solutions, or to solutions at all,
for that matter.
Our gauge choice implies that $\Psi$ obeys $Q\Psi +\Psi^2 = 0$,
while the equation of motion is $(Q - \eta_0)\Psi + \Psi^2 = 0$.
Hence, $\Psi$ is a solution iff $\eta_0\Psi = 0$.
This is exactly the equation of motion of the non-polynomial theory.

The representation~(\ref{PhiForPsi}) does not fix $\Phi$ completely. The redundancy of
$\Phi$ is realised as a new gauge symmetry,
\begin{equation}
\delta e^\Phi = Q \hat \La_0 e^\Phi\,. 
\end{equation}
There is also the residual gauge symmetry, which in terms of $\Phi$, takes the form,
\begin{equation}
\delta e^\Phi = e^\Phi \eta_0 \hat \La_1 \,. 
\end{equation}
These two transformations are identical to those of the non-polynomial theory.

Consider now the action, which using our gauge choice can be written as,
\begin{equation}
S=-\oint \cO\Big(\frac{1}{6}\Psi Q \Psi-\frac{1}{2}\Psi \eta_0 \Psi\Big),
\end{equation}
where the string field on which the action depends is $\Phi$ of~(\ref{PhiForPsi}).
Define now,
\begin{equation}
\Psi(\al)=e^{-\al \Phi}Q e^{\al \Phi}.
\end{equation}
The action of $\Psi(\al)$ equals,
\begin{equation}
S(\al)=-\oint \cO\Big(\frac{1}{6}\Psi(\al) Q \Psi(\al)-\frac{1}{2}\Psi(\al) \eta_0 \Psi(\al)\Big).
\end{equation}
This expression obeys the initial condition $S(0)=0$ and the first order differential equation,
\begin{equation}
\frac{dS(\al)}{d\al}=\oint (Q\cO) A \eta_0 \Phi\,.
\end{equation}
It can be seen that the analogous expressions for the non-polynomial theory is,
\begin{equation}
\frac{dS(\al)}{d\al} = \oint A \eta_0 \Phi\,.
\end{equation}
In light of~(\ref{OandXp}), these expressions match even off-shell. We conclude that the gauge fixed action of the
democratic theory equals (even off-shell) that of the non-polynomial theory. Together with
the previous observations we conclude that the non-polynomial theory is a
partially gauge fixed version of the democratic theory.

The Ramond sector can also be included by using a Ramond string field
$\Xi$, with $pic(\Xi)=\frac{1}{2}$. The partially-gauge-fixed action takes the form,
\begin{equation}
S=S_{NS}-\oint \cO_0 \Big(\frac{1}{2}\eta_0 \Xi Q\eta_0 \Xi
   + e^{-\Phi}Q e^\Phi (\eta_0 \Xi)^2\Big),
\end{equation}
where $\cO_0=P=-c \xi \partial \xi e^{-2\phi}$ is the residual mid-point insertion.
It can be shown that this action results in the desired equations of motion and gauge symmetry.

\subsection{Gauge fixing at $pic(\Psi)=-1$}

Now, the condition~(\ref{etaPsi0}) becomes inconsistent at the non-linear level.
Its modification to a consistent condition is just the $\Ztwo$-dual\cite{Berkovits:1994vy} of the condition that
gives the modified theory at $pic(\Psi)=0$,
\begin{equation}
\Psi=e^\Phi \eta_0 e^{-\Phi}\,.
\end{equation}
The resulting theory is, again, the non-polynomial theory, only in its $\Ztwo$-dual form.
The condition~(\ref{QPsi0}), on the other hand, is now consistent at the non-linear level.
The resulting theory reads,
\begin{equation}
S=\oint \cO_2 \Big(\frac{1}{2}\Psi\eta_0 \Psi+\frac{1}{3}\Psi^3 \Big)\,.
\end{equation}

Alternatively, one can write the form of this theory in the dual Hilbert space, in which it is defined
in light of the gauge condition
(we use the new integration symbol to represent CFT expectation values in the dual Hilbert space),
\begin{equation}
S=\dint X_2 \Big(\frac{1}{2}\Psi\eta_0 \Psi+\frac{1}{3}\Psi^3 \Big)
\end{equation}
That this theory is the $\Ztwo$ dual of the modified theory is now manifest.
This is not Witten's theory, but a new consistent NS string field theory.
We also showed explicitly that this theory supports analytical
solutions analogous to the ones that exist in the modified theory, i.e.,
marginal and vacuum solutions. We further showed that these solutions have the desired properties.

\section{Conclusions and outlook}
\label{sec:conc}

The democratic theory is the only superstring field theory to date that can be used to represent
all possible sectors of the open superstring. Furthermore, all sectors are unified into a single string
field that lives in the maximal possible space consistent with the D-brane system
(large Hilbert space with arbitrary picture and ghost numbers at all consistent internal and external
Chan-Paton factors).

We showed that in its NS sector the democratic theory can be consistently reduced to the modified
as well as to the non-polynomial string field theory. In the latter case the Ramond sector can also
be included and the result is a full RNS string field theory fixed at given picture numbers.

In much of this construction we implicitly assumed some properties of the space of string fields:
\begin{itemize}
\item We assumed that string fields cannot have mid-point insertions.
\item We relied on a decay as a function of picture number of string fields.
\item In order to derive the classical master equation we had to assumed that the induced inner
      product is non-degenerate.
\end{itemize}
These assumptions do not differ in essence from the usual implicit assumptions regarding string field spaces.
In this sense our theory is not better nor is it worse than any other string field theory in the market.
Nonetheless, we feel that this state of affairs is unsatisfactory.
We should uncover the fundamental
mathematical structures that could be used in order to define string field spaces.

We mentioned that the democratic theory obeys the classical master equation. The proof is almost
trivial, since the algebraic structures used by the theory are identical in form to those of
the bosonic theory. However, we also know that a complete theory should obey not only the classical,
but also the quantum master equation. For this equation to hold without modifying the theory,
the expression $\Delta S$ of the BV formalism should vanish. It is known that this expression is
singular in the bosonic case~\cite{Thorn:1988hm}. Can it be regularised in our case? It seems
that in order to answer this question we should first improve our understanding regarding the
space of string fields mentioned above and regarding the realisation of supersymmetry.

We characterized the democratic theory as the only string field theory that can describe the
superstring. Nonetheless, it is not quite clear how to describe space-time supersymmetry using
this theory. In the RNS formalism, on which we rely, space-time supersymmetry is an emerging concept.
It is realised using the zero-momentum integrated massless Ramond vertex operators and as such it carries
half-integer picture numbers. It is neither conceptually clear which pictures should be taken,
nor is it clear in practice whether a combination of these operators at different pictures exist
such that the resulting currents commute with the $\cO$ mid-point insertion. A better understanding of
this issue might help in further restricting the form of $\cO$ and it might also introduce some
practical tools for string field theoretical research.

We already mentioned that gauge fixing to a given picture number is not the natural thing to do
from the perspective of the democratic theory. It would be interesting to find other gauge fixings
that span many (all?) picture numbers, while appropriately constraining the string fields
(other than the trivial ones that are related to the picture-fixed ones by conjugation).
A particularly useful approach towards such a construction would consist of using the BV formulation of
the democratic theory with an appropriate gauge fixing fermion. One can also try to use the
BV form of the theory when reducing it to the non-polynomial theory using the gauge fixing that we
described. This might enable us to overcome the hard problems one encounters when
attempting a direct BV treatment of the non-polynomial theory, as described by Shingo Torii
in this conference.

We showed that two different gauge fixings lead from the democratic
theory to the modified and the non-polynomial theories. Thus, these two theories should have
the same space of solutions. Nonetheless, it was recently claimed that there is a discrepancy
between the spaces of solutions of these two theories~\cite{Erler:2010pr}.
What could be the meaning of that? One resolution would be that the assumptions of~\cite{Erler:2010pr}
regarding the form of the solutions are too strict. Another option would be that the
gauge fixings considered here are not fully adequate. That might be the case, since we showed
that they are adequate only at the linearised level with respect to the string fields, i.e.,
at the level of vertex operators. Note, however, that even the old Siegel gauge is only well established
at this level. While we cannot fully justify the gauge fixings, we believe that if
a problem really exists, it is a problem with only a small class of states for any of the
gauge-fixed theories and at any rate is not a problem with the democratic theory itself.

Other than the two RNS string field theories to which we showed that the democratic theory can
be reduced, there is also another superstring field theory, namely the pure-spinor string field theory
of~\cite{Berkovits:2005bt}. It would be very interesting to see whether a relation between these two
formulations can be established. Indeed, relations between the RNS and pure-spinor formulations
exist~\cite{Berkovits:2001us} and some elements of this correspondence were actually used in the derivation
of the democratic theory. Moreover, both theories are cubic and are defined using a single mid-point
insertion, which possess no kernel. Nonetheless, this task is far from being trivial, since the
string fields should be related also off-shell, where the structure of the two formalisms
seems quite different. Presumably, a modification of the pure-spinor formalism is needed before
such a relation could be established. We hope that the idea of relating the pure-spinor and the
democratic string field theories could help in the search for a modified pure-spinor formulation.

Finally, one might hope that the ideas behind the democratic theory could be generalised to the
cases of the heterotic and the closed strings. If so, this could lead to fully RNS
heterotic and the closed string field theories, which so far have not been obtained,
mainly due to problems related to picture numbers. There are some essential differences
between the open string case and the other cases. It might be possible to overcome some of the
problems using ideas from~\cite{Saroja:1992vw}. We currently study this research direction.

\newpage
\section*{Acknowledgements}

It it a pleasure to thank the organisers of ``SFT2010 -- the third
international conference on string field theory and related topics'', held at
the YITP Kyoto, for inviting me to present this work.
I would further like to thank the organisers for hospitality and for a
very enjoyable and stimulating environment. 
I would also like to acknowledge discussions on and around the subject
presented here with the following participants of the conference:
Yuri Aisaka, Irina Aref'eva, Loriano Bonora, Ted Erler, Taichiro Kugo, Carlo Maccaferri, Yoji Michishita,
Sanefumi Moriyama, Masaki Murata, Yuji Okawa, Martin Schnabl and Shingo Torii.

My research is supported by an Outgoing International Marie Curie
Fellowship of the European Community. The views presented in this work are
those of the author and do not necessarily reflect those of the European
Community.

%

\end{document}